\documentclass[preprint,12pt]{elsarticle}

\usepackage{amssymb}

\journal{Nuclear Physics B}

\begin{document}

\begin{frontmatter}



\title{Three Dimensional Charged Black Hole Inspired by Noncommutative Geometry}


\author{Alexis Larra\~{n}aga}

\author{Juan Manuel Tejeiro}
\address{Observatorio Astronomico Nacional. Facultad de Ciencias. Universidad Nacional de Colombia}

\begin{abstract}
We find a new charged black hole in three-dimensional anti-de Sitter
space using an anisotropic perfect fluid inspired by the noncommutative
black hole as the source of matter and a gaussian distribution of
electric charge. We deduce the thermodynamical quantities of this
black hole and compare them with those of a charged BTZ solution. 

\end{abstract}


\end{frontmatter}


\section{Introduction}

The theoretical discovery of radiating black holes \cite{hawking}
disclosed the first physically relevant window on the mysteries of
quantum gravity. The string/black hole correspondence principle \cite{sisskind}
suggests that in this extreme regime stringy effects cannot be neglected.
One of the most interesting outcomes of string theory is that target
spacetime coordinates become noncommuting operators on a D-brane \cite{witten}.
Thus, string-brane coupling has put in evidence the necessity of spacetime
quantization. Now, the noncommutativity of spacetime can be encoded
in the commutator

\begin{equation}
\left[x^{\mu},x^{\nu}\right]=i\theta^{\mu\nu},\end{equation}
 where $\theta^{\mu\nu}$ is an anti-symmetric matrix which determines
the fundamental cell discretization of spacetime much in the same
way as the Planck constant $\hbar$ discretizes the phase space. This
proposal provides a black hole with a minimum scale $\sqrt{\theta}$
known as the noncommutative black hole \cite{key-1,key-2,nicolini,key-5,key-7},
whose commutative limit is the Schwarzschild metric. The thermodynamics
and evaporation process of the noncommutative black hole has been
studied in \cite{myung1} while the entropy issue is discussed in
\cite{banarjee1,banarjee2} and Hawking radiation in \cite{nozari}.
Charged noncommutative black hole has been studied in \cite{chargednc,chargednc2}
and recently, a noncommutative threedimensional black hole whose commutative
limit is the non-rotating BTZ solution was studied in \cite{myung2},
while the threedimensional rotating counter part was deduced in \cite{tejlar}.

In this paper, we construct a new charged black hole in $AdS_{3}$
spacetime using an anisotropic perfect fluid inspired by the 4D noncommutative
black holeas the source of matter and considering a gaussian distribution
of electric charge. The resulting solution exhibits two horizons that
degenerate into one in the extremal case. We compare the thermodynamics
of this noncommutaive black hole with that of the charged BTZ solution
\cite{BTZ1,BTZ2}.

\section{Derivation of the Charged Solution in Three Dimensions}

It has been shown \cite{key-1,key-2} that the noncommutativity eliminates
point-like structures in favor of smeared objects in flat space-time.
A way of implementing the effect of smearing is a substitution rule:
in three dimensions, the Dirac delta function $\delta^{3D}\left(r\right)$
is replaced by a Gaussian distribution with minimal width $\sqrt{\theta}$,

\begin{equation}
\rho\left(r\right)=\frac{M}{4\pi\theta}e^{-r^{2}/4\theta}\end{equation}

giving a mass distribution with the form

\begin{eqnarray}
m\left(r\right) & = & 2\pi\int_{0}^{r}r'\rho\left(r'\right)dr'=M\left(1-e^{-r^{2}/4\theta}\right).\end{eqnarray}

As coordinate non-commutativity is a property of the spacetime fabric
itself, and not of its matter content, the same smearing effect is
expected to operate on electric charges \cite{chargednc,chargednc2}.
Thus, a point-charge $Q$ is spread into a minimal width gaussian
charge cloud according to

\begin{equation}
\rho_{e}\left(r\right)=\frac{Q}{4\pi\theta}e^{-r^{2}/4\theta}.\end{equation}
For a static, circular symmetric charge distribution, the current
density $J_{\mu}$ is non-vanishing only along the time direction,
i.e.

\begin{equation}
J^{\mu}=\left(\rho_{e},0,0\right).\label{eq:source}\end{equation}

In order to find a black hole solution in $AdS_{3}$ space-time, we
introduce the Einstein and Maxwell equations,

\begin{eqnarray}
R_{\mu\nu}-\frac{1}{2}g_{\mu\nu}R & = & 8\pi\left(T_{\mu\nu}^{matter}+T_{\mu\nu}^{electr}\right)+\frac{1}{\ell^{2}}g_{\mu\nu}\label{eq:EFE}\\
\frac{1}{\sqrt{\left|g\right|}}\partial_{\mu}\left(\sqrt{\left|g\right|}F^{\mu\nu}\right) & = & J^{\nu},\label{eq:ME}\end{eqnarray}
where $\ell$ is related with the cosmological constant by 

\begin{equation}
\Lambda=-\frac{1}{\ell^{2}}.\end{equation}
The energy-momentum tensor for matter will take the anisotropic form

\begin{equation}
\left(T_{\nu}^{\mu}\right)^{matter}=\mbox{diag}\left(-\rho,p_{r},p_{\bot}\right).\end{equation}

In order to completely define this tensor, we rely on the covariant
conservation condition $T_{\quad,\nu}^{\mu\nu}=0$. This gives the
source as an anisotropic fluid of density $\rho$, radial pressure

\begin{equation}
p_{r}=-\rho\end{equation}

and tangential pressure

\begin{equation}
p_{\bot}=-\rho-r\partial_{r}\rho.\end{equation}

The electromagnetic energy-momentum tensor $T_{\mu\nu}^{electr}$
is defined in terms of $F_{\mu\nu}$ as

\begin{equation}
T_{\mu\nu}^{electr}=-\frac{2}{\pi}\left(F_{\mu\sigma}F_{\nu}^{\;\sigma}-\frac{1}{4}g_{\mu\nu}F_{\rho\sigma}F^{\rho\sigma}\right).\end{equation}
By solving the Maxwell equations (\ref{eq:ME}) with source (\ref{eq:source}),
we obtain the electric field

\begin{eqnarray}
E\left(r\right) & = & \frac{1}{r}\int_{0}^{r}r'\rho_{e}\left(r'\right)dr'=\frac{Q}{2\pi r}\left(1-e^{-r^{2}/4\theta}\right).\end{eqnarray}

Using the static, circular symmetric line element

\begin{equation}
ds^{2}=-f\left(r\right)dt^{2}+f^{-1}\left(r\right)dr^{2}+r^{2}d\varphi^{2},\label{eq:solucionBTZ}\end{equation}

the Eintein's field equations (\ref{eq:EFE}) are written as

\begin{eqnarray}
\frac{1}{r}\frac{df}{dr} & = & -16\pi\rho-\frac{1}{2}E^{2}+\frac{2}{\ell^{2}}\\
\frac{d^{2}f}{dr^{2}} & = & 16\pi\rho_{\bot}+\frac{1}{2}E^{2}+\frac{2}{\ell^{2}}.\end{eqnarray}

Solving the above equations, we find 
\begin{equation}
f\left(r\right)=-8M\left(1-e^{-r^{2}/4\theta}\right)+\frac{r^{2}}{\ell^{2}}-\frac{Q^{2}}{8\pi^{2}}\left[\ln\left|r\right|+\frac{1}{2}\mbox{Ei}\left(-\frac{r^{2}}{2\theta}\right)-\mbox{Ei}\left(-\frac{r^{2}}{4\theta}\right)\right],\label{eq:functionf}\end{equation}
where $\mbox{Ei}\left(z\right)$ represents the exponential integral
function,

\begin{equation}
\mbox{Ei}\left(z\right)=-\int_{-z}^{\infty}\frac{e^{-t}}{t}dt.\end{equation}
Note that when $\frac{r^{2}}{4\theta}\rightarrow\infty$, either for
considering a large black hole $\left(r\rightarrow\infty\right)$
or for considering the commutative limit $\left(\theta\rightarrow0\right)$,
we obtain the charged BTZ solution,

\begin{equation}
f^{BTZ}\left(r\right)=-8M+\frac{r^{2}}{\ell^{2}}-\frac{Q^{2}}{8\pi^{2}}\ln\left|r\right|.\end{equation}

The line element (\ref{eq:solucionBTZ}), (\ref{eq:functionf}) describes
the geometry of a noncommutative black hole with event horizons given
by the condition

\begin{equation}
f\left(r_{\pm}\right)=-8M\left(1-e^{-r_{\pm}^{2}/4\theta}\right)+\frac{r_{\pm}^{2}}{\ell^{2}}-\frac{Q^{2}}{8\pi^{2}}\left[\ln\left|r_{\pm}\right|+\frac{1}{2}\mbox{Ei}\left(-\frac{r_{\pm}^{2}}{2\theta}\right)-\mbox{Ei}\left(-\frac{r_{\pm}^{2}}{4\theta}\right)\right]=0.\end{equation}
This equation cannot be solved in closed form. However, by plotting
$f\left(r\right)$ one can read intersections with the $r$-axis and
determine numerically the existence of horizon(s) and their radii.
Fig. 1 shows that, instead of a single event horizon, there are different
possibilities for this black hole,

1. Two distinct horizons for $M>M_{o}$

2. One degenerate horizon (extremal black hole) for $M=M_{o}$

3. No horizon for $M<M_{o}$.

\begin{center}
\begin{minipage}[t]{1\columnwidth}%
\begin{center}
\includegraphics[scale=0.6]{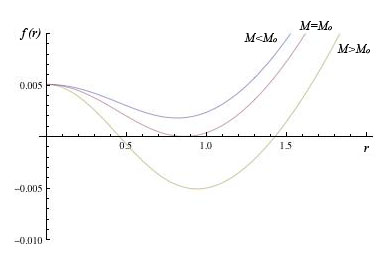}
\par\end{center}

{\footnotesize Fig. 1. Metric function $f$ as a function of $r$.
We have taken the values $\theta=0.1$, $\ell=10$ and $Q=1$. The
minimum mass is $M_{o}\approx0.00127$.}%
\end{minipage} 
\par\end{center}

In view of this results, there can be no black hole if the original
mass is less than the lower limit mass $M_{o}$. The horizon of the
extremal black hole is determined by the conditions $f=0$ and $\partial_{r}f=0,$
which gives

\begin{equation}
\frac{4\left(1-e^{r_{o}^{2}/4\theta}\right)\theta+r^{2}}{\frac{1}{2}Ei\left(-\frac{r_{o}^{2}}{2\theta}\right)-Ei\left(-\frac{r_{o}^{2}}{4\theta}\right)+\ln\left|r_{o}\right|+\frac{2\theta}{r_{o}^{2}}\left(3+e^{-r_{o}^{2}/2\theta}-3e^{-r_{o}^{2}/4\theta}-e^{r_{o}^{2}/4\theta}\right)}=\frac{Q^{2}\ell^{2}}{8\pi^{2}}\end{equation}
and then, the mass of the extremal black hole can be written as 

\begin{equation}
M_{o}=\frac{\frac{2r_{o}^{2}}{\ell^{2}\theta}-\frac{Q^{2}}{8\pi^{2}\theta}\left(1+e^{-r^{2}/2\theta}-2e^{-r^{2}/4\theta}\right)}{4e^{-r_{o}^{2}/4\theta}}.\end{equation}

\section{Thermodynamics}

The Hawking temperature of the noncommutative black hole is

\begin{equation}
T_{H}=\frac{1}{4\pi}\left.\partial_{r}f\right|_{r_{+}}\end{equation}

\begin{equation}
T_{H}=\frac{r_{+}}{2\pi\ell^{2}}\left[1-\frac{2M_{H}\ell^{2}}{\theta}e^{-r_{+}^{2}/4\theta}-\frac{Q^{2}\ell^{2}}{16\pi^{2}r_{+}^{2}}\left(1+e^{-r_{+}^{2}/2\theta}-2e^{-r_{+}^{2}/4\theta}\right)\right],\end{equation}
where

\begin{equation}
M_{H}=\frac{r_{+}^{2}}{8\ell^{2}\left(1-e^{-r_{+}^{2}/4\theta}\right)}-\frac{Q^{2}}{64\pi^{2}\left(1-e^{-r_{+}^{2}/4\theta}\right)}\left[\ln\left|r_{+}\right|+\frac{1}{2}\mbox{Ei}\left(-\frac{r_{+}^{2}}{2\theta}\right)-\mbox{Ei}\left(-\frac{r_{+}^{2}}{4\theta}\right)\right].\end{equation}

For large black holes, i.e. $\frac{r_{+}^{2}}{4\theta}>>0$, one recovers
the temperature of the rotating BTZ black hole,

\begin{equation}
T_{H}^{BTZ}=\frac{r_{+}}{2\pi\ell^{2}}\left[1-\frac{Q^{2}\ell^{2}}{64\pi^{2}r_{+}^{2}}\right].\end{equation}

As shown in the Fig. 2, the Hawking temperature is a monotonically
increasing function of the horizon radius for large black holes. For
large black holes there is no difference with the charged BTZ solution.

\begin{center}
\begin{minipage}[t]{1\columnwidth}%
\begin{center}
\includegraphics[scale=0.6]{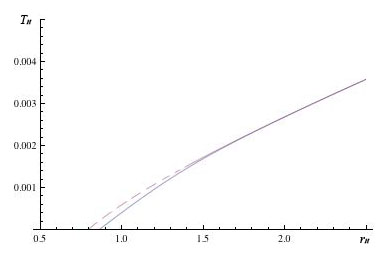}
\par\end{center}

{\footnotesize Fig. 2. Hawking temperature versus $r_{H}$. The solid
line represents the temperature for the noncommutative black hole
with $\theta=0.1$. There is no difference ith charged BTZ solution
(dashed line) for large $r_{H}$. In both cases we use the values
$\ell=10$ and $Q=1$. }%
\end{minipage}
\par\end{center}

The first law of thermodynamics for a charged black hole reads

\begin{equation}
dM=T_{H}dS+\Phi dQ,\end{equation}

where the electrostatic potential is given by

\begin{equation}
\Phi=\left(\frac{\partial M}{\partial Q}\right)_{r_{+}}=-\frac{Q}{32\pi^{2}\left(1-e^{-r_{+}^{2}/4\theta}\right)}\left[\ln\left|r_{+}\right|+\frac{1}{2}\mbox{Ei}\left(-\frac{r_{+}^{2}}{2\theta}\right)-\mbox{Ei}\left(-\frac{r_{+}^{2}}{4\theta}\right)\right].\end{equation}

We calculate the entropy as

\begin{equation}
S=\int_{r_{o}}^{r_{+}}\frac{1}{T_{H}}dM\end{equation}

which gives

\begin{equation}
S=\frac{\pi}{2}\int_{r_{o}}^{r_{+}}\left(\frac{1}{1-e^{-\xi^{2}/4\theta}}\right)d\xi.\end{equation}

The entropy as a function of $r_{+}$ is depicted in Fig. 3. Note
that, in the large black hole limit, the entropy function corresponds
to the Bekenstein-Hawking entropy (area law), $S_{BH}=\frac{\pi r_{+}}{2}$.

\begin{center}
\begin{minipage}[t]{1\columnwidth}%
\begin{center}
\includegraphics[scale=0.6]{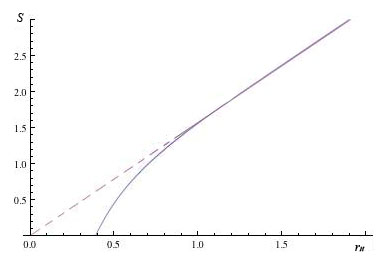}
\par\end{center}

{\footnotesize Fig. 3. Entropy versus $r_{H}$. The solid line represents
the entropy of the noncommutative black hole with $\theta=0.1$. The
dashed line represents the entropy of the charged BTZ black hole. }%
\end{minipage}
\par\end{center}

\section{Conclusion}

We construct a noncommutative electric charged black hole in $AdS_{3}$
spacetime using an anisotropic perfect fluid inspired by the 4D noncommutative
black hole and a gaussian distribution of electric charge. The resulting
solution two horizons that degenerate into one in the extreme case.
We compare the thermodynamics of this black hole with that of a charged
BTZ black hole. The Hawking temperature and entropy of large noncommutative
charged black hole approach those of charged BTZ solution. 




\bibliographystyle{elsarticle-num}

\end{document}